\documentclass[prb,twocolumn,showpacs,preprintnumbers,amsmath,amssymb]{revtex4}

\usepackage{graphicx}
\usepackage{amsmath}
\usepackage{epstopdf}
\usepackage{amsbsy}
\usepackage{dcolumn}
\usepackage{bm}

\newcommand{\jvinter}{J_{\rm v}^{\rm inter}}
\newcommand{\jbinter}{J_{\rm b}^{\rm inter}}
\newcommand{\jvintra}{J_{\rm v}^{\rm intra}}
\newcommand{\jbintra}{J_{\rm b}^{\rm intra}}

\newcommand{\jb}{J_{\rm b}}
\newcommand{\jv}{J_{\rm v}}

\begin{document}

\title{Magnetic field-induced soft mode in spin-gapped high-T$_c$ superconductors}

\author{Brian M. Andersen$^1$, Olav F. Sylju\aa sen$^2$, and Per Hedeg\aa rd$^1$}

\affiliation{$^1$Niels Bohr Institute, University of Copenhagen, Universitetsparken 5, DK-2100 Copenhagen, Denmark\\
$^2$Department of Physics, University of Oslo, P.O. Box 1048 Blindern, N-0316 Oslo, Norway}

\date{\today}

\begin{abstract}

We present an explanation of the dynamical in-gap spin mode in La$_{2-x}$Sr$_x$CuO$_4$ (LSCO) induced by an applied magnetic field $H$ as recently observed by J. Chang {\it et al.} \cite{chang09} Our model consists of a phenomenological spin-only Hamiltonian, and the softening of the spin mode is caused by vortex pinning of dynamical stripe fluctuations which we model by a local ordering of the exchange interactions. The spin gap vanishes experimentally around $H=7$T which in our scenario corresponds to the field required for overlapping vortex regions.

\end{abstract}

\pacs{74.25.Ha,74.72.-h,75.10.Jm,75.40.Gb} 

\maketitle

The cuprate superconductors (SC) arise from doping an antiferromagnetic (AF) Mott insulator. At half-filling the spin spectrum of La$_2$CuO$_4$ is quantitatively described by a spin-1/2 Heisenberg model.\cite{coldea} At finite doping, however, the nature of the magnetic fluctuations and their importance for SC remains controversial. At present, the so-called "hour-glass" dispersion observed in inelastic neutron response appears universal whereas details of the low-energy spin fluctuations vary between the compounds.\cite{tranquadareview} In the optimal and overdoped regimes an interplay between magnetism and SC has been revealed by the opening of a spin gap which scales with the SC transition temperature $T_c$. This is in contrast to the underdoped regime where LSCO and Bi$_2$Sr$_2$CaCu$_2$O$_{8+\delta}$ (BSCCO) are know to exhibit spin freezing well into the SC dome,\cite{julien,wakimoto,panagopoulos,suzuki} whereas YBa$_2$Cu$_3$O$_{6+x}$ (YBCO) reveals a static signal only at very low doping.\cite{haug} 

In the quest of obtaining a better understanding of these materials, the effect of an applied magnetic field $H$ has been extensively used, also in neutron scattering experiments. 
In the underdoped regime a magnetic field applied perpendicular to the CuO$_2$ planes enhances static incommensurate (IC) stripe order, which exist at $H=0$ presumably due to impurities.\cite{andersen07,andersen08} This enhanced signal at the IC positions $q^*$ [quartet of peaks near $(\pi,\pi)$] has been seen both in underdoped La$_{2-x}$Sr$_x$CuO$_4$, \cite{katano,lake02,chang08} La$_2$CuO$_{4+y}$, \cite{khaykovich02} and very recently also in YBa$_2$Cu$_3$O$_{6.45}$.\cite{haug} A similar enhancement of stripe order can be obtained at $H=0$ by the use of impurity substitution.\cite{ChNiedermayer:1998,HKimura:2003,yslee04,fujita} 
Furthermore, experiments have shown that spin-gapped samples can transition from a SC phase to a coexisting SC and IC stripe ordered phase by use of a magnetic field.\cite{khaykovich05,chang08} 

In LSCO, static order is absent for doping levels beyond approximately $x\sim 0.13$.\cite{julien} The inelastic spin excitations are, however, still characterized at low energy by the same IC wavevectors but a spin gap of order $\sim 3-7$ meV develops at $T<T_c$.\cite{mason92,yamada95,lake99,wakimoto04} The $H$-dependence of the low-energy inelastic neutron response has also been subject of intense experimental investigations. Lake {\it et al.}\cite{lake01} reported a softening of the spin mode in LSCO ($x=0.163$) revealed by an in-gap mode observed with $H=7.5$T compared to a fully spin-gapped spectrum for $H=0$. Similar results have been obtained at larger doping levels.\cite{gilardi04,tranquada04} More recently detailed inelastic neutron scattering experiments studied the $H$-dependence of the magnetic spin gap in slightly underdoped LSCO ($x=0.145$).\cite{chang09} At this doping, a critical field of $H_c=7$T is required to tune the system from a SC state into a phase with coexisting SC and long-range IC stripe order. At applied fields $0<H<7$T the spin gap is diminished and an in-gap spin mode is observed.\cite{lake01,chang09} This transition is very reminiscent of the dynamic neutron response seen e.g. by Kimura {\it et al.}\cite{HKimura:2003} in Zn-doped LSCO ($x=0.15$). 

The presence of SC regions appear important for the existence of enhanced magnetic response at these relatively low applied fields $H\sim 0-10T$; only when vortices can act as additional pinning centers for nucleation of stripe regions do they lead to an enhanced magnetic response. This agrees qualitatively with: 1) the fact that the enhanced signal is seen at $T<T_c$, and 2) an absent [a negligible] magnetic field effect in non-SC [weakly SC] samples.\cite{matsuda02,wakimoto03,tranquada08} It is not necessary for the vortices to form an ordered lattice which also appear absent at $x=1/8$ in LSCO.\cite{chang08} Theoretically, the existence of AF order induced by vortices was first discussed within the SO(5) theory of the cuprates.\cite{sczhang,arovas,abh} Later several models studied how vortices may nucleate magnetic regions due to a general competition between SC and stripe order.\cite{demler,kivelson,chen,zhu,takigawa,ahb,linda}

One way to model the stripe phase is in terms of coupled spin ladders.\cite{kruger,vojta,uhrig,yao,andsyl,konik} In such spin-only models, the charge carriers are assumed important only for renormalizing the exchange couplings between localized spinful regions, and the Hamiltonian is of the Heisenberg form. Clearly this approach is phenomenological, and should be considered an approximate effective model. Nevertheless, this approach has been very successful is describing e.g. the universal "hour-glass" magnetic dispersion. 


Here, motivated by the recent experimental findings of Chang {\it et al.},\cite{chang09} we study theoretically the effect of a magnetic field on the low-energy (gapped) spin fluctuations. 
By extending the coupled spin ladder approach to include the effect of vortices, we find a field-induced mode inside the gap similar to experiment. The vortices are modeled by local regions of exchange couplings $J_v$ different from the bulk $J_b$. Thus, the effective Hamiltonian reads 
\begin{equation}
H=\sum_{b\langle ij \rangle} J_b {\mathbf{S}}_i \cdot {\mathbf{S}}_j + \sum_{v\langle ij \rangle} J_v {\mathbf{S}}_i \cdot {\mathbf{S}}_j, 
\end{equation}
where $\langle ij \rangle$ denote neighboring spins, and $b$ and $v$ refer to effective exchange couplings far away from (in the vicinity of) the vortices $\jb=\{\jbintra,\jbinter\}$ ($\jv=\{\jvintra,\jvinter\}$). For simplicity, we assume that the intra-ladder exchange couplings are unaffected by the vortices $\jbintra=\jvintra=J$, and that only the inter-ladder couplings may differ between $b$ bulk and $v$ vortex regions.

\begin{figure}[t]
\includegraphics[clip=true,width=.7\columnwidth]{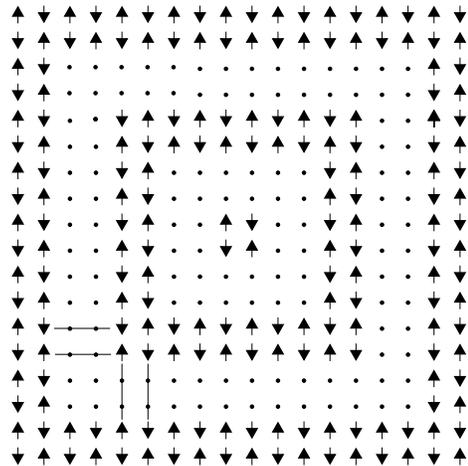}
\caption{Arrangement of stripes around a vortex core. The arrows indicate spins and show an ordered configuration according to the signs of the couplings. The dots are nonmagnetic sites. The AF interaction between neighboring spins on the same ring is $\jvintra$ while spins on different rings are coupled by a ferromagnetic interaction $\jvinter$ which are indicated, for a few spins, as straight lines in the lower left corner.} \label{moellebrett}
\end{figure}

Self-consistent microscopic studies of the Hubbard model have shown that, depending on input parameters, impurities and vortices can induce uni-directional stripes, approximately rotational invariant stripes (ARI stripes), or checkerboard patterns in the surrounding magnetization and hole density.\cite{chen,zhu,andersenpriv,zha} Furthermore, the induced spin density is found to be modulated with a period close to eight lattice spacings.\cite{chen,zhu} Consistent with these results we model the vortex regions by ARI stripes and arrive at a picture of the spins near the vortices as shown in Fig.~\ref{moellebrett}. Clearly, this spin arrangement is idealized, but we believe it gives a reasonable description of the structure near the vortices. For example, as shown in Ref. \onlinecite{ahb}, stripe arrangements similar to Fig.~\ref{moellebrett} agree with the checkerboard local density of states modulations observed near vortices in BSCCO.\cite{hoffman} We stress, however, that the particular choice of induced order shown in Fig.~\ref{moellebrett} is not important for the following discussion; similar results can be obtained from uni-directional stripes. Checkerboards, on the other hand, do not appear to be consistent with experiment.\cite{abh} The important point is that vortices in underdoped cuprates nucleate stripe order in a surrounding halo characterized by a new length scale $\xi_{IC}$, which lies between the SC coherence length $\xi$ and penetration depth $\lambda$ for these materials, $\xi<\xi_{IC}<\lambda$. Typical spatial extent of the vortex regions used in our simulations is taken to be $\sim 50\times 50$ lattice sites consistent with estimates of the correlation length $\xi_{IC} \sim 100\AA$.\cite{chang09,lake02} 

The value of the inter-ladder exchange coupling $\jvinter$ is restricted by the fact that for high magnetic fields the vortex regions overlap and long-range IC order is observed,\cite{lake02} implying that $|\jvinter|$ should exceed the critical value $J_{\rm c}^{\rm inter}\approx -0.4J$ (for 2-leg ladders\cite{andsyl}) for appearance of long-range order. Thus, in the following we simply fix $\jvinter=-0.6 J$, i.e. well below $J_{\rm c}^{\rm inter}$. On the other hand, the {\em absence} of long-range spin order at $H=0$ implies that the bulk inter-ladder exchange couplings $\jbinter$ are weaker than this critical value. While it is possible to obtain the results reported here by assigning a small value to $|\jbinter|$, we prefer instead to view $\jbinter$ as an effective coupling obtained by randomly diluting bonds of strength $\jvinter$. Thus, some of the $\jbinter$ are $0$ while others are $-0.6J$, mimicking the disordered nature of stripes in the bulk $b$.\cite{andersen07,robertson,delmaestro,kaul,alvarez,ah} By contrast, we assume that the primary effect of a magnetic field is to introduce vortex regions of {\em ordered} stripes where all the inter-ladder bonds have strength $\jvinter=-0.6J$. The motivation for this effective "exchange-ordering" nature of $H$ comes from recent quantum oscillation measurements which indicate that an external magnetic field cause exactly this kind of ordering of the stripes resulting in a severe Fermi surface reconstruction.\cite{doiron}

In order to find the spin excitation spectra we simulated Heisenberg spin systems as depicted in Fig.~\ref{moellebrett} at a temperature $T=0.01J$ using the quantum Monte Carlo stochastic series expansion technique with directed-loop updates.\cite{SS} This technique yields high quality results for the imaginary-time correlation function which is continued to real frequencies using the Average Spectrum Method,\cite{ASM} where the average over all possible spectra is taken weighted by how well each spectrum fits the data. This approach performs at least as well as MaxEnt methods for high quality imaginary-time data.\cite{stocon} 

\begin{figure}[t]
\includegraphics[clip=true,height=.75\columnwidth,width=1.0\columnwidth]{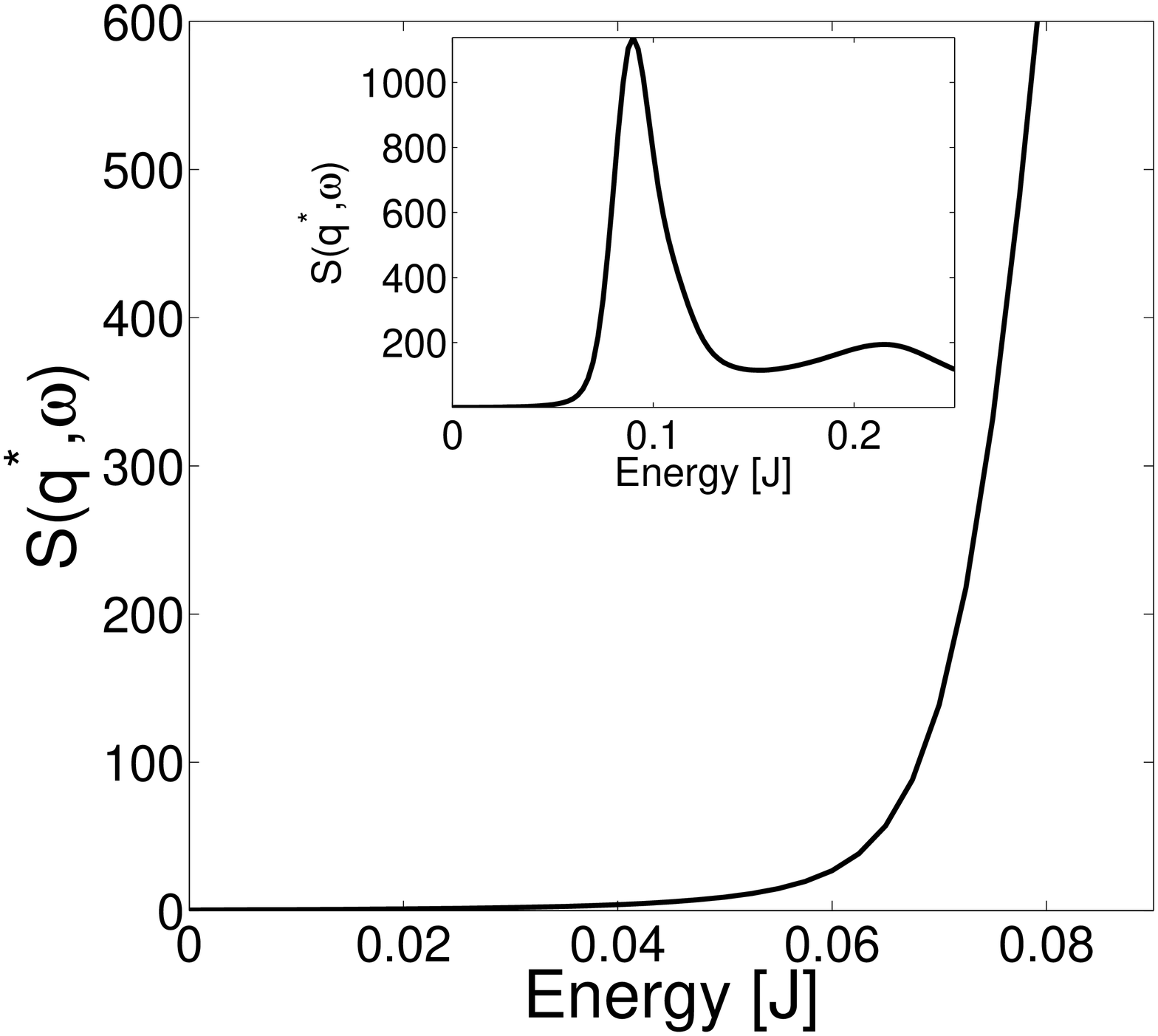}
\caption{Magnetic structure factor $S_b(q^*,\omega)$ at the IC wavevector $q^*=(3\pi/4,\pi)$  as a function of energy $\omega$. This spin spectrum is obtained from a disordered set of coupled 2-leg spin ladders mimicking the neutron response from disordered stripes at $H=0$. The inset shows the same quantity for a larger energy range.} \label{fig1}
\end{figure}

The bulk part $S_b(q^*,\omega)$ and the vortex part $S_v(q^*,\omega)$ of the structure factor were simulated separately. 
For the bulk we diluted the arrangement of $\jbinter$ bonds randomly prior to performing the simulations. The structure factor $S_b(q^*,\omega)$ was extracted for each disorder realization and the average was taken. With an interladder bond dilution fraction of $30\%$ we found that $S_b(q^*,\omega)$ has a sharp peak around $\omega=0.1J$ as well as a broader high energy peak at $\omega=0.2J$ as seen from the inset in Fig.~\ref{fig1}. This two-peak structure at the IC position $q^*$ is consistent with inelastic neutron scattering measurements on spin-gapped LSCO.\cite{vignolle} Without a microscopic model including realistic disorder concentrations it is hard to estimate the correct degree of bond disorder within the present spin-only approach. However, the position of the low energy peak depends on the amount of bond dilution, less dilution moves the peak down in energy, and we have simply chosen $30\%$ in order to reproduce a spin gap of order $10\%$ of $J$ as reported in Ref. \onlinecite{chang09}.

\begin{figure}[t]
\includegraphics[clip=true,height=.75\columnwidth,width=1.0\columnwidth]{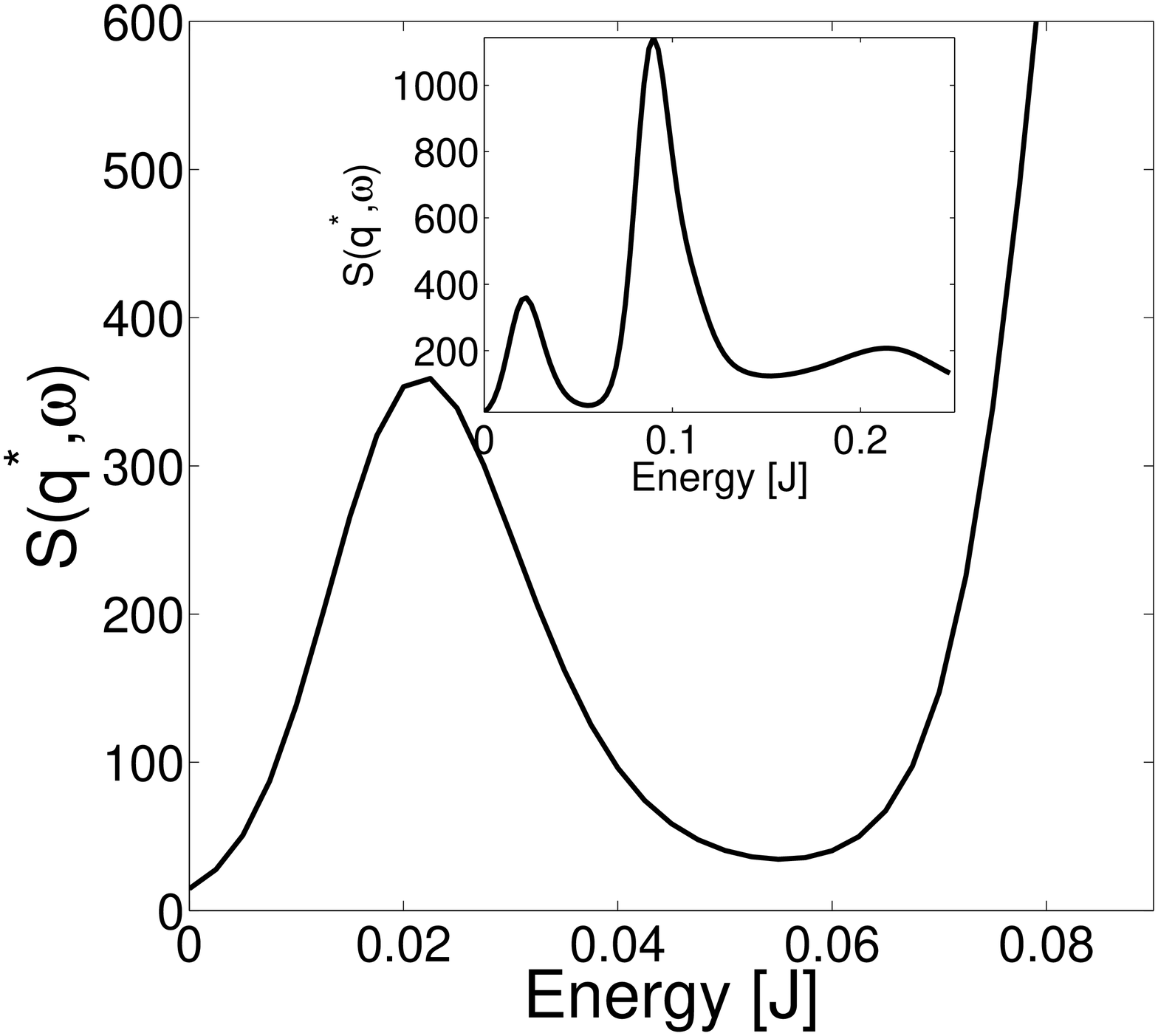}
\caption{Total magnetic structure factor $S(q^*,\omega)$ at the IC wavevector $q^*=(3\pi/4,\pi)$ as a function of energy $\omega$ including the effects of an applied magnetic field.The inset shows the same quantity for a larger energy range.} \label{fig2}
\end{figure}

For the vortex regions we use the non-diluted configuration to obtain $S_v(q^*,\omega)$ which has a structure similar to Fig.~\ref{fig1} except that the low-energy peak has moved further down in energy. For $\jvinter=-0.6J$ the lowest peak is roughly at $\omega=0.02J$ (see Fig.~\ref{fig2}). This corresponds to $\omega\sim 2$meV in agreement with Ref. \onlinecite{chang09} using recent estimates for the exchange coupling $J\sim 100$meV in LSCO ($x=0.14$).\cite{vignolle} The peak position is dependent on both the finite size of the vortex region and the value of $\jvinter$; as expected smaller vortex regions (or smaller value of $|\jvinter|$) moves the peak upwards in energy. Taking into account that for $H=2.5T$ (as used in Ref. \onlinecite{chang09}) the bulk region contributes roughly 15 times more to the total $S(q^*,\omega)$ than the vortex regions, we compose the total structure factor $S(q^*,\omega)=(15 S_b(q^*,\omega)+S_v(q^*,\omega))/16$ shown in Fig.~\ref{fig2}. As seen, $H$ induces an in-gap mode similar to the experimental results of Refs. \onlinecite{chang09} and \onlinecite{lake01}.  

At low magnetic fields [$H \lesssim 3$T in LSCO ($x=0.145$)], an increase of $H$ enhances the vortex density, but the size of each vortex region is presumably unchanged resulting in a fixed energy of the in-gap spin mode. In this field range the spectral weight of the in-gap mode should increase as a consequence of an increased weighting of the total structure factor by the denser amount of vortex regions. By contrast when vortex regions overlap, the mode will rapidly move to zero energy becoming a true Bragg signal. Vortex regions of size $\sim 100\AA$ will start overlapping at $H\sim 6$T in agreement with the closing of the spin gap found in Ref. \onlinecite{chang09} near this field strength. Theoretically, the generation of a Bragg peak happens if $\jvinter$ is larger than the critical value $J_{\rm c}^{\rm inter}$ needed for long-range stripe order. By contrast, if $|\jvinter|<|J_{\rm c}^{\rm inter}|$, a finite energy peak can still show up at finite field, but it will not move to zero as the vortex regions start overlapping. 

In conclusion we have proposed an explanation for the soft magnetic mode observed inside the spin gap of LSCO in terms of magnetic field-induced stripe ordered vortex regions. At zero applied magnetic field $H=0$ the CuO$_2$ planes are modeled as disordered coupled spin ladders known to reproduce the magnetic "hour-glass" dispersion seen by experiments.  At finite field $H\neq 0$ vortices are simulated by regions of ordered exchange couplings. Our calculation shows that stripe pinning by vortices gives a consistent picture of the in-gap spin mode observed in recent inelastic neutron scattering measurements. 

The authors acknowledge valuable discussions with J. Chang, N. B. Christensen, and K. Lefmann. B.~M.~A. was supported by the Villum
Kann Rasmussen foundation. The computer calculations were carried out using resources provided by
the NOTUR project of the Norwegian Research Council.


\begin{thebibliography}{00}

\bibitem{chang09} J. Chang, N. B. Christensen, Ch. Niedermayer, K. Lefmann, H. M. R\o nnow, D. F. McMorrow, A. Schneidewind, P. Link, A. Hiess, M. 
Boehm, R. Mottl, S. Pailh\'{e}s, N. Momono, M. Oda, M. Ido, and J. Mesot, arXiv:0902.1191v1.
%
\bibitem{coldea} R. Coldea, S. M. Hayden, G. Aeppli, T. G. Perring, C. D. Frost, T. E. Mason, S.-W. Cheong, and Z. Fisk, Phys. Rev. Lett. {\bf 86}, 5377 (2001). 
%
\bibitem{tranquadareview} For a review, see J. M. Tranquada, in {\it Handbook of High-Temperature Superconductivity Theory and Experiment}, edited by J. R. Schrieffer (Springer, New York,  2007).
%
\bibitem{julien} M.-H. Julien, Physica B {\bf 329-333}, 693  (2003).
%
\bibitem{wakimoto} S. Wakimoto, R. J. Birgeneau, Y. S. Lee, and G. Shirane, Phys. Rev. B {\bf 63}, 172501 (2001).
%
\bibitem{panagopoulos} C. Panagopoulos  J. L. Tallon, B. D. Rainford, J. R. Cooper, C. A Scott, and T. Xiang, Solid State Comm. {\bf 126}, 47 (2003).
%
\bibitem{suzuki} T. Suzuki, T. Goto, K. Chiba, T. Shinoda, T. Fukase, H. Kimura, K. Yamada, M. Ohashi, and Y. Yamaguchi, Phys. Rev. B {\bf 57}, R3229 (1998). 
%
%
\bibitem{haug} D. Haug, V. Hinkov, A. Suchaneck, D. S. Inosov, N. B. Christensen, Ch. Niedermayer, P. Bourges, Y. Sidis, J. T. Park, A. Ivanov, C. T. Lin, J. Mesot, and B. Keimer, arXiv:0902.3335v1.
%
\bibitem{lake02} B. Lake, H. M. R\o nnow, N. B. Christensen, G. Aeppli, K. Lefmann, D. F. McMorrow, P. Vorderwisch, P. Smeibidl, N. Mangkorntong, T. Sasagawa, M. Nohara, H. Takagi,  and T. E. Mason, Nature (London) {\bf 415}, 299 (2002).
%
\bibitem{andersen07} B. M. Andersen, P. J. Hirschfeld, A. P. Kampf, and M. Schmid, Phys. Rev. Lett. {\bf 99}, 147002 (2007).
%
\bibitem{andersen08} B. M. Andersen and P. J. Hirschfeld, Phys. Rev. Lett. {\bf 100}, 257003 (2008).
%
\bibitem{katano} S. Katano, M. Sato, K. Yamada, T. Suzuki, and T. Fukase, Phys. Rev. B {\bf 62}, R14677 (2000).
%
\bibitem{chang08} J. Chang, Ch. Niedermayer, R. Gilardi, N. B. Christensen, H. M. R\o nnow, D. F. McMorrow, M. Ay, J. Stahn, O. Sobolev, A. Hiess, S. Pailhes, C. Baines, N. Momono, M. Oda, M. Ido, and J. Mesot, Phys. Rev. B {\bf 78}, 104525 (2008).
%
\bibitem{khaykovich02} B. Khaykovich, Y. S. Lee, R. W. Erwin, S.-H. Lee, S. Wakimoto, K. J. Thomas, M. A. Kastner, and R. J. Birgeneau, Phys. Rev. B {\bf 66},
014528 (2002).
%
%
\bibitem{ChNiedermayer:1998} Ch. Niedermayer, C. Bernhard, T. Blasius, A. Golnik, A. Moodenbaugh, and J. I. Budnick, Phys. Rev. Lett. {\bf 80}, 3843 (1998).
%
\bibitem{HKimura:2003} H. Kimura, M. Kofu, Y. Matsumoto, and K. Hirota, Phys. Rev. Lett. {\bf 91}, 067002 (2003).
%
\bibitem{yslee04} Y. S. Lee, F. C. Chou, A. Tewary, M. A. Kastner, S. H. Lee, and R. J. Birgeneau, Phys. Rev. B {\bf 69}, 020502 (2004).
%
\bibitem{fujita} M. Fujita, M. Enoki, S. Iikubo, K. Kudo, N. Kobayashi, and K. Yamada, arXiv:0903.5391v1.
%
\bibitem{khaykovich05} B. Khaykovich, S. Wakimoto, R. J. Birgeneau, M. A. Kastner, Y. S. Lee, P. Smeibidl, P. Vorderwisch, and 
K. Yamada, Phys. Rev. B {\bf 71}, 220508 (2005).
%
%
%
\bibitem{mason92} T. E. Mason, G. Aeppli, and H. A. Mook, Phys. Rev. Lett. {\bf 68}, 1414 (1992).
%
\bibitem{yamada95} K. Yamada, S. Wakimoto, G. Shirane, C. H. Lee, M. A. Kastner, S. Hosoya, M. Greven, Y. Endoh, and R. J. Birgeneau, Phys. Rev. Lett. {\bf 75}, 1626 (1995).
%
\bibitem{lake99} B. Lake, G. Aeppli, T. E. Mason, A. Schr\"{o}der, D. F. McMorrow, K. Lefmann, M. Isshiki, M. Nohara, H. Takagi, and S. M. Hayden, Nature (London) {\bf 400}, 43 (1999).
%
\bibitem{wakimoto04} S. Wakimoto, H. Zhang, K. Yamada, I. Swainson, H. Kim, and R. J. Birgeneau, Phys. Rev. Lett. {\bf 92}, 217004 (2004).
%
\bibitem{lake01} B. Lake, G. Aeppli, K. N. Clausen, D. F. McMorrow, K. Lefmann, N. E. Hussey, N. Mangkorntong, M. Nohara, H. Takagi, T. E. Mason, and A. Schr\"{o}der, Science {\bf 291}, 1759 (2001).
%
\bibitem{gilardi04} R. Gilardi, A. Hiess, N. Momono, M. Oda, M. Ido, and J. Mesot, Europhys. Lett. {\bf 66}, 840 (2004).
%
\bibitem{tranquada04} J. M. Tranquada, C. H. Lee, K. Yamada, Y. S. Lee, L. P. Regnault, and H. M. R\o nnow, Phys. Rev. B {\bf 69}, 174507 (2004).
%
\bibitem{matsuda02} M. Matsuda, M. Fujita, K. Yamada, R. J. Birgeneau, Y. Endoh, and G. Shirane, Phys. Rev. B {\bf 66}, 174508 (2002).
%
\bibitem{wakimoto03} S. Wakimoto, R. J. Birgeneau, Y. Fujimaki, N. Ichikawa, T. Kasuga, Y. J. Kim, K. M. Kojima, S.-H. Lee, H. Niko, J. M. Tranquada, S. Uchida, and M. v. Zimmermann, Phys. Rev. B {\bf 67}, 184419 (2003).
%
\bibitem{tranquada08} J. Wen, Z. Xu, G. Xu, J. M. Tranquada, G. Gu, S. Chang, and H. J. Kang, Phys. Rev. B {\bf 78}, 212506 (2008).
%
\bibitem{sczhang} S.-C. Zhang, Science {\bf 275}, 1089 (1997).
%
\bibitem{arovas} D. P. Arovas, A. J. Berlinsky, C. Kallin, and S.-C. Zhang, Phys. Rev. Lett.  {\bf 79}, 2871 (1997).
%
\bibitem{abh} B. M. Andersen, H. Bruus, and P. Hedeg\aa rd, Phys. Rev. B {\bf 61}, 6298 (2000).
%
%
\bibitem{demler} E. Demler, S. Sachdev, and Y. Zhang, Phys. Rev. Lett. {\bf 87}, 067202 (2001). 
%
\bibitem{kivelson} S. A. Kivelson, D.-H. Lee, E. Fradkin, and V. Oganesyan, Phys. Rev. B {\bf 66}, 144516 (2002).
%
\bibitem{chen} Y. Chen and C. S. Ting, Phys. Rev. B {\bf 65}, 180513 (2002).
%
\bibitem{zhu} J. X. Zhu, I. Martin, and A. R. Bishop, Phys. Rev. Lett. {\bf 89}, 067003 (2002).
%
\bibitem{takigawa} M. Takigawa, M. Ichioka, and K. Machida, Phys. Rev. Lett. {\bf 90}, 047001 (2003).
%
\bibitem{ahb} B. M. Andersen, P. Hedeg\aa rd, and H. Bruus, Phys. Rev. B {\bf 67}, 134528 (2003).
%
\bibitem{linda} L. Udby, B. M. Andersen, and P. Hedeg\aa rd, Phys. Rev. B {\bf 73}, 224510 (2006).
%
\bibitem{kruger} F. Kr\"{u}ger and S. Scheidl, Phys. Rev. B {\bf 67}, 134512 (2003).
%
\bibitem{vojta} M. Vojta and T. Ulbricht, Phys. Rev. Lett. {\bf 93}, 127002 (2004).
%
\bibitem{uhrig} G. S. Uhrig, K. P. Schmidt, and M. Gr\"{u}ninger, Phys. Rev. Lett. {\bf 93}, 267003 (2004).
%
\bibitem{yao} E. W. Carlson, D. X. Yao, and D. K. Campbell, Phys. Rev. B. {\bf 70}, 064505 (2004).
%
\bibitem{andsyl} B. M. Andersen and O. F. Sylju\aa sen, Phys. Rev. B {\bf 75}, 012506 (2007).
%
\bibitem{konik} R. M. Konik, F. H. L. Essler, and A. M. Tsvelik, Phys. Rev. B {\bf 78}, 214509 (2008).
%
\bibitem{andersenpriv} B. M. Andersen and P. Hedeg\aa rd, (unpublished).
%
\bibitem{zha} G.-Q. Zha, L.-F. Zhang, H. Meng, H.-W. Zhao, and S.-P. Zhou, Europhys. Lett. {\bf 84}, 17005 (2008).
%
\bibitem{hoffman} J. E. Hoffman, E. W. Hudson, K. M. Lang, V. Madhavan, H. Eisaki, S. Uchida, and J.C. Davis, Science {\bf 295}, 466 (2002).
%
\bibitem{robertson} J. A. Robertson, S. A. Kivelson, E. Fradkin, A. C. Fang, and A. Kapitulnik, Phys. Rev. B {\bf 74}, 134507 (2006).
%
\bibitem{delmaestro} A. Del Maestro, B. Rosenow, and S. Sachdev, Phys. Rev. B {\bf 74}, 024520 (2006).
%
\bibitem{kaul} M. Vojta, T. Vojta, and R. K. Kaul, Phys. Rev. Lett. {\bf 97}, 097001 (2006).
%
\bibitem{alvarez} G. Alvarez, M. Mayr, A. Moreo, and E. Dagotto, Phys. Rev. B {\bf 71}, 014514 (2005).
%
%
\bibitem{ah} B. M. Andersen and P. Hedeg\aa rd, Phys. Rev. Lett. {\bf 95}, 037002 (2005).
%
\bibitem{doiron} N. Doiron-Leyraud, C. Proust, D. LeBoeuf, J. Levallois, J.-B. Bonnemaison, R. Liang, D. A. Bonn, W. N. Hardy, and L. Taillefer, Nature (London) {\bf 447}, 565 (2007).
%
\bibitem{SS} O. F. Sylju{\aa}sen and A. W. Sandvik, Phys. Rev. E {\bf 66}, 046701 (2002).
%
\bibitem{ASM} S. R. White, in {\em Computer Simulation Studies in Condensed Matter Physics III}, edited by D. P. Landau, K. K. Mon, and H.-B. Schuttler Springer-Verlag, Berlin, 1991.
%
\bibitem{stocon} O. F. Sylju{\aa}sen, Phys. Rev. B {\bf 78}, 174429 (2008).
%
\bibitem{vignolle} B. Vignolle, S. M. Hayden, D. F. McMorrow, H. M. R\o nnow, B. Lake, C. D. Frost, and T. G. Perring, Nature Phys.  {\bf 3}, 163 (2007).

\end{thebibliography}
\end{document}